\documentclass[11pt,a4paper]{article}
\usepackage{amsfonts}
\usepackage{amsthm,amssymb}
\def\bea{\begin{equation}}
\def\eea{\end{equation}}

\def\beq{\begin{equation}}
\def\eeq{\end{equation}}
\def\ba{\beq\new\begin{array}{c}}
\def\ea{\end{array}\eeq}
\def\be{\ba}
\def\ee{\ea}
\def\p{\partial}

\def\f{\frac}

\newcommand{\Tr}{\mathop{\rm Tr}\nolimits}

\newcommand{\pP}[1]{\frac{\partial}{\partial#1}}

\def\l{\label}
\def\r{\ref}
\makeatletter
\newdimen\normalarrayskip 
\newdimen\minarrayskip 
\normalarrayskip\baselineskip
\minarrayskip\jot
\newif\ifold \oldtrue \def\new{\oldfalse}
\def\arraymode{\ifold\relax\else\displaystyle\fi} 
\def\eqnumphantom{\phantom{(\theequation)}} 
\def\@arrayskip{\ifold\baselineskip\z@\lineskip\z@
\else
\baselineskip\minarrayskip\lineskip2\minarrayskip\fi}
\def\@arrayclassz{\ifcase \@lastchclass \@acolampacol \or
\@ampacol \or \or \or \@addamp \or
\@acolampacol \or \@firstampfalse \@acol \fi
\edef\@preamble{\@preamble
\ifcase \@chnum
\hfil$\relax\arraymode\@sharp$\hfil
\or $\relax\arraymode\@sharp$\hfil
\or \hfil$\relax\arraymode\@sharp$\fi}}
\def\@array[#1]#2{\setbox\@arstrutbox=\hbox{\vrule
height\arraystretch \ht\strutbox
depth\arraystretch \dp\strutbox
width\z@}\@mkpream{#2}\edef\@preamble{\halign
\noexpand\@halignto
\bgroup \tabskip\z@ \@arstrut \@preamble \tabskip\z@ \cr}%
\let\@startpbox\@@startpbox \let\@endpbox\@@endpbox
\if #1t\vtop \else \if#1b\vbox \else \vcenter \fi\fi
\bgroup \let\par\relax
\let\@sharp##\let\protect\relax
\@arrayskip\@preamble}
\def\eqnarray{\stepcounter{equation}%
\let\@currentlabel=\theequation
\global\@eqnswtrue
\global\@eqcnt\z@
\tabskip\@centering
\let\\=\@eqncr
$$%
\halign to \displaywidth\bgroup
\eqnumphantom\@eqnsel\hskip\@centering
$\displaystyle \tabskip\z@ {##}$%
\global\@eqcnt\@ne \hskip 2\arraycolsep
$\displaystyle\arraymode{##}$\hfil
\global\@eqcnt\tw@ \hskip 2\arraycolsep
$\displaystyle\tabskip\z@{##}$\hfil
\tabskip\@centering
&{##}\tabskip\z@\cr}
\begingroup\ifx\undefined\newsymbol \else\def\input#1 {\endgroup}\fi

\date{\today}
\begin{document}

\setcounter{footnote}{1}
\def\thefootnote{\fnsymbol{footnote}}
\begin{center}
\hfill ITEP/TH-26/02\\
\hfill hep-th/0205261\\
\vspace{0.3in}
{\Large\bf Givental formula in terms of Virasoro operators}
\end{center}
\centerline{{\large A. Alexandrov}\footnote
{ITEP, Moscow, Russia; e-mail: alex@gate.itep.ru}}
\setcounter{footnote}{0}
\def\thefootnote{\arabic{footnote}}
\bigskip
\abstract{\footnotesize
We present a conjecture that the universal enveloping algebra of
differential operators $\frac{\p}{\p t_k}$ over $\mathbb{C}$
coincides in the origin with the universal enveloping algebra of the (Borel
subalgebra of) Virasoro generators from the Kontsevich model. Thus,
we can decompose any (pseudo)differential operator to a combination of
the Virasoro operators. Using this decomposition we present the r.h.s. of the Givental formula \cite{Giv}
as a constant part of the differential operator we introduce.
In the case of $\mathbb{CP}^1$ studied in \cite{Song:2001ww}, the l.h.s.
of the Givental formula is a unit, which imposes certain constraints on this
differential operator. We explicitly check that these constraints are correct
up to $O(q^4)$. We also propose a conjecture of factorization modulo Hirota
equation of the differential operator introduced and check this conjecture with the same accuracy.
}

\begin{center}
\rule{5cm}{1pt}
\end{center}

\bigskip

\section{Introduction}
It is well known that $\mathcal{N}=2$ supersymmetric sigma-model with target space $M$,
defined on the genus $g$ Riemann surface, through
twisting, leads to the topological field theory \cite{Witten:1988xj}. For $D=2$ sigma-model, the
correlation functions of such topological field theory are known to be some intersection
numbers on the target space, and the fields of this theory, which we call primary
fields, are in one-to-one correspondence with the
elements of $H^*(M)$ (see, for example, \cite{Witten:1989ig}).
Thus, for any genus $g$, the free energy $F_M^g(s)$ of the theory, as a
function of couplings $s$ to primary fields (the space of such couplings
is called small phase space) is a generating function for such intersection
numbers on the target space. The genus zero free energy  $F^0_M(s)$ is described by
associativity -- WDVV equations \cite{Witten:1989ig}\cite{WDVV}.

Coupling of the topological field theory to the topological gravity leads to the
so called topological string theory. The genus $g$ free energy $\mathcal{F}^g_M(t)$ of this theory,
which depends on infinitely many
coupling constants $t$ \footnote{We will denote coordinates on the big phase space as
$t$, while on the small phase space, subspace of the big one, as $s$.}
(couplings to gravitational descendants),
describes intersection theory on the compactified moduli spaces of
genus $g$ punctured Riemann surfaces. Combining $\mathcal{F}^g_M(t)$ for all $g$ into the
sum and exponentiating
$\tau_M(t)=\exp{\sum_{g=0}\hbar^{g-1}\mathcal{F}_M^g(t)}$, one gets a general
object which depends on the manifold and is the function of infinitely many coupling
constants. We call it $\tau$-function \cite{Morozov:1995pb}.
This function can be considered from many different points of
view, which lead to different conditions on $\tau_M(t)$
(\cite{Witten:1989ig}\cite{Kontsevich:qz}\cite{Eguchi:1994in}\cite{Eguchi:1996tg})
(Virasoro constraints, dilaton and divisor equations). However, in general,
they do not seem to be restrictive enough to determine $\tau_M(t)$ completely.

In fact, there are at least two examples of the manifold $M$ when the function $\tau_M$ for topological
field theory with the target space $M$ turns out to be the $\tau$-function of integrable
hierarchy.
The first example is the point. Its function
$\tau_\bullet$, which corresponds to the pure topological
gravity \cite{Witten:1989ig}, is the $\tau$-function for
the KdV hierarchy \cite{Fukuma:1990yk},\cite{Douglas:1989dd}. It can be
realized as the matrix integral (famous Kontsevich model\cite{Kontsevich:ti})
\be
\tau_\bullet(T)=\frac{\int dH e^{-1/3\Tr{H^3}+\Tr{M^2H}}}{\int dH e^{-\Tr{MH^2}}},
\ee
over $N \times N$ Hermitian matrix, the coupling
constants, i.e. the times from the integrable point of view, are defined as follows
\be
T_k=-\frac{1}{k}\Tr{M^{-k}}.
\ee
Invariance of this matrix integral with
respect to the change of variables \cite{Marshakov:1991nv}, namely,
\be
H\rightarrow H+\epsilon_p,
\ee
with $\epsilon_p=M^{2p}$ leads to the
Virasoro constraints $L_n\tau_\bullet(T_k)=0,~n\geq-1$ (see, for
instance, \cite{Witten:1991mn},\cite{Marshakov:1991fu},\cite{Kharchev:1991cy}) with
\footnote{In our paper, we use another normalization and numeration of times, that is,
\be
T_{2m+1}=\frac{\Gamma(\frac{3}{2})}{\Gamma(m+\frac{3}{2})}t_m,
\ee
which comes from enumerative geometry.
}
\be
L_n=\frac{1}{2}\sum_{k~odd}kT_k\frac{\p}{\p T_{k+2n}}+\frac{1}{4}\sum_{{a+b=2n}\atop{a,b~odd
~and>0}}\frac{\p^2}{\p T_a\p
T_b}+\frac{T_1^2}{4}\delta_{n,-1}+\frac{1}{16}\delta_{n,0}-\frac{\p}{\p
T_{3+2n}}, ~~~n\geq -1.
\l{vvv}
\ee
These Virasoro operators form the Borel subalgebra of the Virasoro algebra,
\be
[L_n,L_m]=(n-m)L_{n+m}.
\ee

The second example, $M=\mathbb{CP}^1$,
is conjectured to correspond to the Toda hierarchy \cite{Eguchi:1994in}\cite{Getzler:2001}.
The $\tau$-function for it can also be
represented as some matrix integral. By
analogous consideration, this representation leads to the Borel subalgebra of some
other Virasoro algebra
which annihilates the $\tau$-function. Therefore, one
could expect that the function $\tau_M$ associated with
any manifold, is the $\tau$-function for some integrable
hierarchy. This is on of the reasons why we generally call it $\tau$-function.

For several particular manifolds there exist some explicit formulas for
$F_M^g$. The example of $M=\mathbb{CP}^1$ is, in a sense, the only foreseeable:
its whole free energy vanishes on the small phase space, except genus zero
and genus one, which are, correspondingly,
\footnote{
Hereafter, we denote as $s_P$, $s_Q$ and $s_R$ the three coupling
constants, corresponding to the cohomology  classes 1, $\omega$ and
$\omega^2$, with $\omega$ being the 2-form $\omega\in H^2(M)$.}
\be
F_{\mathbb{CP}^1}^0=\frac{1}{2}(s_P)^2s_Q+e^{s_Q}
\l{cp10}
\ee
and
\be
F_{\mathbb{CP}^1}^1=-s_Q/24.
\l{cp11}
\ee

For $\mathbb{CP}^2$ the genus zero free energy is of the form
\be
F_{\mathbb{CP}^2}^0=\frac{1}{2}s_P(s_Q)^2+\frac{1}{2}(s_P)^2s_R+\sum_{d=1}N_d^{(0)}
\frac{(s_R)^{3d-1}}{(3d-1)!}\exp{(s_Qd)},
\l{ser1}
\ee
and for the expansion coefficients $N_d^{(0)}$ there is the recursive relation \cite{Kontsevich:qz}
\footnote{The numbers
$N_d^{(0)}$ and $N_d^{(1)}$ are numbers of degree $d$ rational (elliptic)
curves passing through $3d-1$ (respectively $3d$) points.}
\be
N_d^{(0)}=(3d-4)!\sum_{l+k=d}\frac{N_l^{(0)}N_k^{(0)}}{(3l-1)!(3k-1)!}k^2l[3kl+l-2k].
\l{N0}
\ee
For the genus one free energy
\be
F_{\mathbb{CP}^2}^1=\frac{-s_Q}{8}+\sum_{d=1}N_d^{(1)}\frac{(s_R)^{3d}}{(3d)!}\exp{(ds_Q)},
\l{ser2} \ee
analogous recursion relation was derived in
\cite{Eguchi:1997bg}, using Virasoro constraints for $\mathbb{CP}^n$ and the
topological recursion relations (see also \cite{Getzler:1996})
\be
N_n^{(1)}=N_n^{(0)}\frac{1}{72}n(n-1)(n-2)+\sum_{k+l=n}(3n-1)!\frac{N_k^{(0)}}{(3k-1)!}
\frac{N_l^{(1)}}{(3l)!}\frac{l}{9}(3k^2-2k).
\l{N1}
\ee
However, in general, there are no
expressions for higher genera similar to (\r{N0}),(\r{N1}),
and only a few first terms in the series like (\r{ser1}),(\r{ser2})
are found.

The problem is to construct a regular procedure for finding $F_g^M(s)$. In this paper, we make
a modest contribution to solving this problem.
It is based on the remarkable formula which was proposed in \cite{Giv} with the help of the localization technique, and
connects the Kontsevich $\tau$-function ($\tau_\bullet$, function
for the point on the big phase space) with the $\tau$-function for some (for instance,
projective) manifold $M$ on the small phase space, with the genus zero and one contributions
omitted
\footnote{ This formula
was proposed for the manifolds such that $F^0_M(s)$ determines a semi-simple Frobenius structure
on the space $H^*(M,Q)$ \cite{Giv}.}
\be
e^{\sum_{g>1}\hbar^{g-1}F^g_M(s)}=
\Big[e^{D_M(s)}\prod_j
\tau_\bullet\big(\hbar\Delta_j;t_0^{(j)},t_1^{(j)},\ldots\big)\Big]_{t_k^{(i)}= T_k^i(s)}.
\l{MF}
\ee
Here $s_i$ are coordinates on the small phase space for the
manifold $M$, indices $i,j$ run from 1 to $K=dim(H^*(M))$, and
$D_M(s)$ is the bilinear pseudodifferential
\footnote{We call it pseudodifferential, since it contains infinite number of derivatives.
For the sake of brevity, we sometimes call it differential operator.}
operator acting on $t_k^{(i)}$
\be
D_M(s)=\frac{\hbar}{2}\sum_{k,l=0}^\infty\sum_{i,j}V_{kl}^{ij}(s)\Delta_i^{1/2}(s)\Delta_j^{1/2}(s)
\p_{t_k^{(i)}}\p_{t_l^{(j)}},
\ee
$V_{kl}^{ij}(s)$, $\Delta_i(s)$ and $T_k^i(s)$ being some functions
which can be presented in terms of
the genus zero potential $F^0_M(s)$, more precisely on its third derivatives (\cite{Giv},\cite{Song:2001ww}).

Formula (\r{MF}) was studied in \cite{Song:2001ww} for the first
nontrivial example of the manifold $M$, $\mathbb{CP}^1$.

Since, as we know from the explicit form of the free energy on the small
phase space for $\mathbb{CP}^1$ the l.h.s. of (\r{MF} is unity, formula
($\r{MF}$) can be considered as an equation imposed on
the Kontsevich $\tau$-function. We call it Song equation.
\footnote{This equation is bilinear,
and it can be presented in the form, reminiscent of the Hirota equation. The obvious
difference is that the Hirota equation is true for any values of times, whereas
the equation (\r{MF}) should be considered
on the one-parameter family of points, since, in this case, $V$, $\Delta$ and $T$
depend only on $s_Q$, the coupling with the 2-form on $\mathbb{CP}^1$.}
In \cite{Song:2001ww} all the necessary
coefficients $\Delta,V$ and $T$ were calculated, and (\r{MF}) was checked
extensively using the expansion in $\hbar$ with the accuracy of $O(\hbar^4)$. This
calculation involves the explicit form of $\tau_\bullet(t)$ as a series in
times.

In our paper, we reformulate the problem of calculation of the r.h.s. in
(\r{MF}) for given $V$ and $T$ realizing some
particular differential operator $D'_M$ easily connected with $D_M$ (see
(\r{DD})) via Virasoro operators.

Namely, the shift operator

\be
\exp{\sum_{i=1}^K\sum_{k=2}^\infty T^i_k(s)\p_{t^{(i)}_k}}
\ee
obviously commutes with $D_M(s)$, and for the operator

\be
D'_M(s)=D_M(s)+\sum_{i=1}^K\sum_{k=2}^\infty T^i_k(s)\p_{t^{(i)}_k}
\l{DD}
\ee
we have the equality

\be
\Big[e^{D_M(s)}\prod_j
\tau_\bullet\big(\hbar\Delta_j;t_0^{(j)},t_1^{(j)},\ldots\big)\Big]_{t_k^{(i)}=T^i_k(s)}=
\Big[e^{D'_M(s)}\prod_j
\tau_\bullet\big(\hbar\Delta_j;t_0^{(j)},t_1^{(j)},\ldots\big)\Big]_{t_k^{(i)}=0}.
\l{MF2}
\ee

The general idea of the paper is to present the operator $e^{D'_M}$ in the origin of the space
of times $t^{(i)}_k$ in terms of the Virasoro operators (this mean, that we consider r.h.s. modulo term,
proportional to the positive powers of the times $t$ )(\r{vvv})
\be
e^{D'_M}\Big|_{t=0}=\sum_{[l_1],[l_2],\ldots,[l_K]}^\infty P^M_{[l_1],[l_2],\ldots[l_K]}(s)
L_{[l_1]}\otimes L_{[l_2]}\otimes\ldots\otimes L_{[l_K]}\Big|_{t=0},
\ee
Here $L_{[l]}$ are products of the Virasoro
operators with the nonnegative multiplicities $m_i$, given by multiindices $[l]$:
for $[l]=(m_{-1},m_0,\ldots,m_r)$ we define
\be
L_{[l]}:=L_r^{m_r}\ldots L_0^{m_0}L_{-1}^{m_{-1}},
\ee
and, if all multiplicities $m_i$ vanish,
\be
L_{[0]}=1,
\l{id}
\ee
where we denote [0]:=(0,0,\ldots,0). Since all the Virasoro operators of the Borel subalgebra annihilate
the Kontsevich $\tau$-function, the action of the
operator presented in this way on the $\tau_\bullet$'s at the point $t=0$ is trivial,
that is, the only nonvanishing contribution to the l.h.s. of
(\r{MF}) is given by the constant term proportional to (\r{id}):
\be
e^{\sum_{g>1}\hbar^{g-1}F^g_M(s)}=P^M_{[0],[0],\ldots[0]}(s).
\ee

In particular, the Song equation in terms of $D'$ is

\be
\exp\Big[D'_{\mathbb{CP}^1}\Big]
\tau(\hbar,x)\tau(\hbar,y)\Big|_{x_n=y_n=0}=1.
\l{Gc}
\ee
In the case of $\mathbb{CP}^1$, one can expand the operator $D'_{\mathbb{CP}^1}$ in the series
in $q:=\exp{(-\frac{s_Q}{2})}$

\be
e^{D'_{\mathbb{CP}^1}}=1+\sum_{n=1}^\infty q^n
\sum_{[l_1],[l_2]}P^{\mathbb{CP}^1(n)}_{[l_1],[l_2]}(\hbar)L_{[l_1]}\otimes
L_{[l_2]}\Big|_{t=0},
\l{gag}
\ee
Since for $\mathbb{CP}^1$ the l.h.s. of (\r{MF}) is equal to 1, the Song equation
is equivalent to the condition
\be
P^{\mathbb{CP}^1(n)}_{[0],[0]}=0,~~n\geq1.
\l{hren}
\ee
In the paper, we check this condition with the accuracy of $O(q^4)$.

The paper contains two conjectures. The first one which states
that any differential operator in times $t$ with constant coefficients can be
presented as a combination of the Virasoro operators with some multiplicities at the origin
helps to deal with the Givental formula. The second conjecture, whose general structure is
still unknown, in particular case of $M=\mathbb{CP}^1$ claims that
the operator $D'_{\mathbb{CP}^1}$ (\r{gag})
can be factorized into the sum of two parts, which are tensor squares of differential operators,
modulo Hirota equations. Both this conjectures are checked perturbatively in
the example of $\mathbb{CP}^1$ with the accuracy of $O(q^4)$.

The paper is organized as follows: in section 2 we represent the main results of
\cite{Song:2001ww} and explicitly present the Virasoro and Hirota constraints
which we use in our calculations.
Sections 3 and 4 are devoted to our two conjectures, while section 5
contains some concluding remarks.

\section{Main Ingredients}
In the paper \cite{Song:2001ww} it was shown that, denoting $t^{(+)}_n=x_n$ and
$t^{(-)}_n=y_n$, one has the following differential operator
\be
D_{\mathbb{CP}^1}:=
\frac{\hbar}{2}\sum_{k,l\geq0}(V^{++}_{kl}(\p_{x_k}\p_{x_l}
+\p_{y_k}\p_{y_l})+2i(-1)^{l-1}V^{+-}_{kl}\p_{x_k}\p_{y_l}),
\l{Op}
\ee
and introducing $\alpha$ being $c$-numbers
\be
\alpha_{m+1}=-\frac{(2m+1)^2}{8(m+1)}\alpha_m,~~~\alpha_0=1.
\ee
we can represent times\footnote{Since $T_n^-=(-1)^{n-1}T_n^+$, we should define only one set of $T$'s.} $T^+$
\be
T^+_n =
\left\{
\begin{array}{cl}
0, &n=0,1,\\
-\frac{\alpha_{n-1}}{2^{n-1}}\exp\Big[\frac{-(n-1)s_Q}{2}\Big], &n\geq 2.
\end{array}
\right.
\l{SpecT}
\ee
In the case of $\mathbb{CP}$ we have two times on the small phase space but
third derivatives of the free energy depend only on one coordinate, $s_Q$.
This mean that all components of our construction depends on this one
coordinate. $\alpha$ being $c$-numbers
\be
\alpha_{m+1}=-\frac{(2m+1)^2}{8(m+1)}\alpha_m,~~~\alpha_0=1.
\ee
Coefficients $V^{++}$ and $V^{+-}$ in (\r{Op}) can be presented as bilinear combinations of $T_k^+$
similar to those
in the paper \footnote{The proper formula has a slight mistake therein.}\cite{Song:2001ww}:
\be
V^{++}_{kl}=\sum_{n=0}^{k-1}\frac{(-1)^n(4(l+n+1)(k-n)-1)}{(2l+2n+1)(2k-2n-1)}T^+_{l+n+2}T^+_{k-n+1}+
(-1)^{k+1}\frac{T^+_{l+k+2}}{2l+2k+1},\\
V^{+-}_{kl}=i(-1)^l\Big[\sum_{n=0}^{k-1}\frac{2(l+2n+1-k)}{(2l+2n+1)(2k-2n-1)}T^+_{l+n+2}T^+_{k-n+1}+
2(l+k+1)\frac{T^+_{l+k+2}}{2l+2k+1}\Big].
\l{koef}
\ee
Taking into account that, in (\r{SpecT})
$T_0^+=T_1^+=0$, one can explicitly substitute
the Kontsevich $\tau$-function to ($\r{MF}$) in the case $M=\mathbb{CP}^1$, and check the equality,
expanding the operator and $\tau$-functions in the $\hbar$-series
\be
\exp\Big[D_{\mathbb{CP}^1}\Big]
\tau_\bullet(\hbar,x)\tau_\bullet(\hbar,y)\Big|_{x_n=y_n=T^+_n}=:1+\sum_{n=1}^\infty
a_n\hbar^n.
\l{hser}
\ee
In this way, one gets an infinite family of equations on the coefficients of
the decomposition of $\tau_\bullet(\hbar,t)$
\be
a_n=0,~~\forall n>0
\l{any}
\ee
In the paper \cite{Song:2001ww}, the first three equations were explicitly checked.

The Virasoro operators corresponding to our normalization
of times, are (see, for example, \cite{Getzler:1998})
\be
L_n =
\left\{
\begin{array}{cl}
\sum_{m=1}^\infty \tilde{t}_m\p_{m-1}+\frac{1}{2\hbar}t_0^2, & n=-1,\\
\sum_{m=0}^\infty (m+\frac{1}{2})\tilde{t}_m\p_{m}+\frac{1}{16}, & n=0,\\
\sum_{m=0}^\infty
\frac{\Gamma(n+m+\frac{3}{2})}{\Gamma(m+\frac{1}{2})}\tilde{t}_m\p_{m+n}+
\frac{\hbar}{2}\sum_{m=0}^{n-1} (-1)^{m+1}\frac{\Gamma(n-m+\frac{1}{2})}{\Gamma(-m-\frac{1}{2})}
\p_{m}\p_{n-m-1}, & n>0,\\
\end{array}
\right.
\l{vir}
\ee
with $\tilde{t}_m:=t_m-\delta_{m,1}$, $\p_m=\pP{t_m}$. They annihilate the Kontsevich
$\tau$-function,
\be
L_n\tau_\bullet(\hbar,t)=0,~n\geq-1.
\ee

It is also known that any KdV $\tau$-function, in particular, the Kontsevich
one, solves the Hirota bilinear equations, the first three
of which being (see, for instance, \cite{Date:1982tj})
\be
(\hbar D_0^4-12D_0D_1)\tau(t)\cdot\tau(t)=0,\\
(-24D_1^2+\hbar D_0^3D_1+\frac{\hbar^2}{12}D_0^6)\tau(t)\cdot\tau(t)=0,\\
(-D_1^2+\frac{\hbar^2}{360}D_0^6+D_0D_2)\tau(t)\cdot\tau(t)=0,
\ee
which we denote for the sake of brevity as
\be
\mathbf{H}_i=0,~i=1,2,3.
\ee
\section{Connection between Song and Virasoro equations}
Consider an arbitrary operator of the form
\be
B:=\p_1^{n_1}\p_2^{n_2}\ldots\p_k^{n_k},~~n_i\geq0,~~\sum_{i=1}^k n_k>0,
\l{ado}
\ee
for some finite $k$. We experimentally check for small $k$ and $n$
\footnote{
We expressed 27 derivatives, which are necessary for our perturbative
calculation (see below), in terms of the Virasoro operators. Using explicit
form of the Virasoro operators we can get some answers for actions of the
differetial operators \l{ado} on the tau-function, not representing them as a combination
of Virasoro operators, for example
\be
\f{\p}{\p T_1}^{n_0}\f{\p}{\p
T_3}^{n_1}\tau_\bullet(T_1,T_3,\ldots)\Big|_{T_1=T_3=\ldots=0}=
\f{1}{3}\Big(\f{3}{2}\Big)^{n_1}\Big(\f{1}{6\hbar}\Big)^{\f{n_0}{3}}\frac{\Gamma(-\frac{n_0}{3})}{\Gamma{(-n_0)}}\f{\Gamma(n_1+\f{1}{3}n_0+\f{1}{24})}
{\Gamma(\f{1}{3}n_0+\f{1}{24})}.
\ee
}
that this operator can be presented as a combination of the Virasoro operators
(\r{vir}) at the origin of the space of times $t$,
\be
B=\sum_{m_{-1},\ldots m_r=0}^\infty c_{m_r,\ldots,m_0,m_{-1}}L^{m_r}_{r}\ldots
L_0^{m_0}L_{-1}^{m_{-1}}\Big|_{t=0}=:\sum_{[l]}c_{[l]}L_{[l]}\Big|_{t=0},
\l{cony}
\ee
for some finite $r$, dependent on $n$ and $k$, where we introduced multiindex
$[l]=(m_{-1},m_0,\ldots,m_r)$. Some of the multiplicities $m_i$, and even all them,
can vanishes.

For example,
\be
\p_0=-L_{-1}\Big|_{t=0},\\
\p_0^2=L_{-1}^2\Big|_{t=0},\\
\p_0^3=-L_{-1}^3\Big|_{t=0}+\frac{1}{\hbar},\\
\p_2=\frac{1}{30}\hbar L_{-1}^2\Big|_{t=0}-\frac{4}{15}L_1\Big|_{t=0}.
\ee
If this is correct for arbitrary $B$, one can present any
pseudodifferential operator as a combination of the Virasoro operators at the
origin of the space of times $t$. Thus, one can present any tensor product of $N$
differential operators as a sum of tensor products of
combinations of the Virasoro operators
\be
\sum_{[l_1],[l_2],\ldots [l_N]}^\infty P_{[l_1],[l_2],\ldots,[l_N]}L_{[l_1]}\otimes
L_{[l_2]}\otimes\ldots\otimes L_{[l_N]}\Big|_{t=0},
\ee
with $P$ being some coefficients.

This representation can be appropriate for our consideration.
In particular, one has
\be
e^{D'_M}=\sum_{[l_1],[l_2],\ldots,[l_K]}^\infty P^M_{[l_1],[l_2],\ldots[l_K]}(s)
L_{[l_1]}\otimes L_{[l_2]}\otimes\ldots\otimes L_{[l_K]}\Big|_{t=0}.
\ee
Here the operator
\be
D'_M(s)=D_M(s)+\sum_{i=1}^K\sum_{k=2}^\infty T^i_k(s)\p_{t^{(i)}_k}
\ee
is the combination of the $D_M$ and the operator, which shifts the times
$t_k^{(i)}=T_k^i$ to zero.

Since all the Virasoro operators of the Borel subalgebra annihilate $\tau_\bullet$,
calculation of
the action of the operator on the product of $\tau_\bullet$'s in the origin is equivalent to
calculating $P_{[0],[0],\ldots,[0]}$. One observes that the l.h.s. of the (\r{MF}) is equal to the
identity component of the decomposition of operator $e^{D'_M}$ in termes of
Virasoro operators
\be
e^{\sum_{g>1}\hbar^{g-1}F^g_M(s)}=P^M_{[0],[0],\ldots[0]}(s).
\ee
In our particular case, moving our times to the origin is equivalent to changing the operator
$D$: for
the operator
\be
D'_{\mathbb{CP}^1}=D_{\mathbb{CP}^1}+\sum_{n=2}^\infty T_n^+(\pP{x_n}+\pP{y_n}),
\ee
the Song equation is
\be
\exp\Big[D'\Big]
\tau_\bullet(\hbar,x)\tau_\bullet(\hbar,y)\Big|_{x=y=0}=1.
\ee
  Now we can check the Song equation perturbatively, not specifying the explicit form
of $\tau_\bullet$.
This is the advantage of our approach, since the approach of the paper \cite{Song:2001ww}
required the explicit form of $\tau_\bullet$.
To study this equation perturbatively, we can expand the operator in the series in
$q$,
\be
\exp\Big[D'_{\mathbb{CP}^1}\Big]=:1+\sum_{n=1}^\infty q^n
\sum_{[l_1],[l_2]}P_{[l_1],[l_2]}^{\mathbb{CP}^1(n)}(\hbar)L_{[l_1]}\otimes L_{[l_2]}\Big|_{t=0},
\ee
and all we must check is that
\be
P_{[0][0]}^{\mathbb{CP}^1(m)}=0~~ \forall m>0.
\l{PPP}
\ee
We check this up to $n=3$.

The explicit expression for the operator
$e^{D'_{\mathbb{CP}^1}}\Big|_{x=y=0}$ acting on $\tau_\bullet$'s looks like
\footnote{This form seems to be convenient, since the
operator $D'_{\mathbb{CP}^1}$ is symmetric in $x$ and $y$}:
\be
e^{D'_{\mathbb{CP}^1}}\tau\otimes\tau\Big|_{x=y=0}=\\
\begin{array}{cl}
&=1+\Big(\big[-\frac{1}{30}L_1\tau+\big(-\frac{7}{120}L_{-1}^2\tau+\frac{1}{8}(L_{-1}\tau)^2\big)\hbar\big]q\\
&+\big[\big(\frac{1}{3600}(L_1\tau)^2+\frac{1}{3600}L_1^2\tau+\frac{1}{504}L_2\tau\big)+
\big(\frac{109}{50400}L_{-1}\tau+\frac{53}{8400}L_{-1}L_{0}\tau\\
&-\frac{1}{240}L_{-1}\tau L_{-1}L_1\tau+\frac{7}{7200}L_{-1}^2L_1\tau
-\frac{7}{240}L_{-1}\tau L_0\tau+\frac{7}{7200}L_{-1}^2\tau L_1\tau\big)\hbar\\
&+\big(\frac{49}{57600}L_{-1}^4\tau-\frac{7}{960}L_{-1}\tau L_{-1}^3\tau+
\frac{499}{57600}(L_{-1}^2\tau)^2\big)\hbar^2\big]q^2+\big[\big(-\frac{97}{453600}L_3\tau\\
&-\frac{1}{648000}L_1^3\tau-\frac{1}{30240}L_1L_2\tau-
\frac{1}{216000}L_1^2\tau L_1\tau-\frac{1}{30240}L_2\tau L_1\tau\big)\\
&+\big(\frac{31}{7200}(L_0\tau)^2+\frac{14951}{3024000}L_{-1}\tau L_1\tau-\frac{53}{504000}L_{-1}L_0\tau L_1\tau-
\frac{53}{504000}L_{-1}L_0L_1\tau\\
&+\frac{1}{4032}L_{-1}\tau L_{-1}L_2\tau-\frac{7}{864000}L_1^2\tau L_{-1}^2\tau+
\frac{1}{28800}L_{-1}\tau L_{-1}L_{1}^2\tau+\frac{1}{28800}(L_{-1}L_1\tau)^2\\
&-\frac{1}{17280}L_{-1}^2L_2\tau-\frac{1}{17280}L_2\tau L_{-1}^2\tau
-\frac{61}{252000}L_0^2\tau-\frac{1319}{1814400}L_{-1}L_1\tau\\
&+\frac{7}{14400}L_{-1}\tau L_0L_1\tau-\frac{7}{864000}L_{-1}^2L_1^2\tau
-\frac{151}{324000}L_0\tau-\frac{7}{432000}L_{-1}^2L_1\tau L_1\tau\\
&+\frac{7}{14400}L_{-1}L_1\tau L_0\tau\big)\hbar+\big(\frac{7}{57600}L_{-1}L_1\tau L_{-1}^3\tau
+\frac{1121}{1728000}L_{-1}\tau L_{-1}^2\tau\\
&-\frac{53}{288000}L_{-1}^3 L_0\tau+\frac{49}{57600}L_{-1}^3\tau L_0\tau
-\frac{4567}{12096000}L_{-1}^3\tau-\frac{1103}{288000}L_{-1}L_0\tau L_{-1}^2\tau\\
&+\frac{7}{57600}L_{-1}\tau L_{-1}^3L_1\tau+\frac{661}{403200}L_{-1}\tau L_{-1}^2L_0\tau
-\frac{499}{1728000}L_{-1}^2L_1\tau L_{-1}^2\tau\\
&
-\frac{49}{3456000}L_{-1}^4L_1\tau-\frac{49}{3456000}L_{-1}^4\tau L_1\tau\big)\hbar^2
+\big(-\frac{6643}{13824000}L_{-1}^4\tau L_{-1}^2\tau\\
&+
\frac{199}{460800}(L_{-1}^3\tau)^2+\frac{49}{460800}L_{-1}\tau L_{-1}^5\tau
-\frac{343}{41472000}L_{-1}^6\tau\big)\hbar^3\big]q^3\Big)\Big|_{t=0}+O(q^4).
\end{array}
\ee
Now, if one uses the explicit expression for the $\tau$-function $\tau_\bullet$,
conditions ($\r{PPP}$) for
$m=3n$ reproduce the conditions $a_n=0$ from (\r{any}). They are only those conditions that
restrict the coefficients $V^{++}$, $V^{+-}$ and $T^+$.

\section{Linear equation from the bilinear one}
The Song equation is quadratic in $\tau_\bullet$.
Thus, if one manage to presents the operator $e^{D'_{\mathbb{CP}^1}}$ as
\be
e^{D'_{\mathbb{CP}^1}}=f(\hbar,q,L)^{\otimes 2}+2\hbar q g(\hbar,q,L)^{\otimes
2},
\l{con}
\ee
the Song equation is equivalent to the system of two linear equations
\be
f(\hbar,q,L)\tau_\bullet\Big|_{t=0}=1,\\
g(\hbar,q,L)\tau_\bullet\Big|_{t=0}=0.
\ee
Here $f$ and $g$ are some operators expressed in terms of the Virasoro operators at
the origin.

It turns out that the conjecture (\r{con}) is slightly incorrect, instead, with the accuracy of
$O(q^4)$, one gets
\be
\Big[e^{D'_{\mathbb{CP}^1}}-f(\hbar,q,L)^{\otimes 2}-2\hbar q g(\hbar,q,L)^{\otimes2}\Big]
\tau\otimes\tau\Big|_{t=0}=\\
\big[\frac{\hbar}{768}q^2\mathbf{H}_1-\big(\frac{11}{12288}\mathbf{H}_2+\frac{21}{1024}\mathbf{H}_3
+\frac{17}{737280}\hbar^2L_{-1}^2\mathbf{H}_1-\frac{1}{46080}\hbar
L_{1}\mathbf{H}_1\big)q^3\big]\Big|_{t=0},
\ee
where $\mathbf{H}$'s are combinations bilinear in $\tau$ which correspond to
the Hirota equations
\be
\begin{array}{cl}
\mathbf{H}_1(t=0)=&\big(\big(2L_{-1}^4\tau-8 L_{-1}\tau
L_{-1}^3\tau+6(L_{-1}^2\tau)^2\big)\hbar\\
&+8L_{-1}\tau-16L_{-1}L_{0}\tau+16L_{-1}
\tau L_{0}\tau\big)\Big|_{t=0}\\
\mathbf{H}_2(t=0)=&\big(32L_0\tau-\frac{64}{3}L_0^2\tau+\frac{64}{3}(L_0\tau)^2
+\big(-2L_{-1}^3\tau+\frac{4}{3}L_{-1}^3L_0\tau\\
&+2L_{-1}\tau L_{-1}^2\tau
-4L_{-1}\tau L_{-1}^2L_0\tau+4L_{-1}L_0\tau L_{-1}^2\tau
-\frac{4}{3}L_{-1}^3\tau L_0\tau\big)\hbar\\
&+\big(-L_{-1}\tau L_{-1}^5\tau+\frac{1}{6}L_{-1}^6\tau+\frac{5}{2}L_{-1}^4\tau
L_{-1}^2\tau-\frac{5}{3}(L_{-1}^3\tau)^2\big)\hbar^2\big)\Big|_{t=0}\\
\mathbf{H}_3(t=0)=&\big(\big(\frac{16}{15}L_0\tau+\frac{8}{15}L_{-1}L_1\tau
-\frac{8}{15}L_{-1}\tau L_1\tau-\frac{8}{9}L_0^2\tau+\frac{8}{9}(L_0\tau)^2\big)\\
&+\big(-\frac{1}{15}L_{-1}^3\tau+\frac{1}{15}L_{-1}\tau L_{-1}^2\tau\big)\hbar
+\big(\frac{1}{180}L_{-1}^6\tau-\frac{1}{30}L_{-1}\tau L_{-1}^5\tau\\
&+\frac{1}{12}L_{-1}^4\tau
L_{-1}^2\tau-\frac{1}{18}(L_{-1}^3\tau)^2\big)\hbar^2\big)\Big|_{t=0}.
\end{array}
\ee

The operators $f$ and $g$ are, respectively,
\be
\begin{array}{cl}
&f(\hbar,q,L)=1+\big(-\frac{1}{60}L_{1}-\frac{7}{240}\hbar L_{-1}^2\big)q
+\big(-\frac{13}{3150}\hbar L_{-1}+\frac{19}{1400}\hbar
L_{-1}L_{0}\\
&+\frac{1}{1008}L_{2}+\frac{7}{14400}\hbar L_{-1}^2L_1
+\frac{1}{7200}L_1^2-\frac{101}{57600}\hbar^2L_{-1}^4\big)q^2+\frac{1}{2}\big(\big(-\frac{97}{453600}L_{3}\\
&-\frac{1}{648000}L_1^3-\frac{1}{30240}L_1L_2\big)
+\big(-\frac{19}{42000}L_{-1}L_0L_1-\frac{59}{63000}L_0^2-\frac{7}{864000}L_{-1}^2L_1^2\\
&-\frac{1}{17280}L_{-1}^2L_2+\frac{937}{162000}L_0
-\frac{311}{28350}L_{-1}L_1\big)\hbar+\big(\frac{101}{3456000}L_{-1}^4L_1\\
&-\frac{187}{756000}L_{-1}^3+\frac{31}{24000}L_{-1}^3L_0\big)\hbar^2
-\frac{793}{41472000}L_{-1}^6\hbar^3\big)q^3+O(q^4)
\end{array}
\ee

\be
\begin{array}{cl}
&g(\hbar,q,L)=\frac{1}{4}L_{-1}+
\big(-\frac{1}{240}L_{-1}L_1+\frac{1}{320}\hbar
L_{-1}^3-\frac{1}{20}L_0)q+\big(\big(\frac{97}{6300}L_1\\
&+\frac{1}{1200}L_0L_1+\frac{1}{4032}L_{-1}L_2
+\frac{1}{28800}L_{-1}L_1^2\big)+\big(-\frac{1}{2400}L_{-1}^2-\frac{47}{33600}L_{-1}^2L_0\\
&-\frac{1}{19200}L_{-1}^3L_1\big)\hbar
-\frac{1}{460800}L_{-1}^5\hbar^2\big)q^2+O(q^3).
\end{array}
\ee
\section{Conclusion}
In the paper, we presented two conjectures, which could help in the "experimental" work
with the Givental formula ($\r{MF}$), and perturbatively checked them for the simplest example of
the manifold $\mathbb{CP}^1$. In spite of the lack
of the complete prove for these
conjectures, and even clear formulation of
the second one, they could be of some use.
One way to use them is to present the differential operator corresponding to the manifold
$M$, completely in terms of the Virasoro operators with the constant
coefficients $P^M_{[l_1],[l_2],\ldots,[l_N]}$ (We should mention, that this is not
equivalent to present $D'_M(s)$ in such form),
\be
e^{D'_M(s)}=\big[\exp{{
\sum_{[l_1],[l_2],\ldots [l_N]}^\infty P^M_{[l_1],[l_2],\ldots,[l_N]}L_{[l_1]}\otimes
L_{[l_2]}\otimes\ldots\otimes L_{[l_N]}}}\big]\Big|_{t=0}.
\ee
This is equivalent to finding the $\tau$-function of the manifold $M$ on the small phase
space, if one additionally knows the genus 1 free energy (the genus zero free energy
explicitly enters the coefficients in the r.h.s.).
We do not know how to do this nonperturbatively, even for $M=\mathbb{CP}^1$.
The other way is to proceed with perturbative calculations of free energies, for example,
in the case of $M=\mathbb{CP}^2$, using formula ($\r{MF}$) and ideas of this paper.

Another interesting way to go is to study the formula analogous to ($\r{MF}$)
such that its l.h.s. contains the free energy (with genus 0 and 1 contributions subtracted)
on the big phase space \cite{Giv}.
It gives almost $\tau_M$, modulo genus zero and genus one
contributions. However, the structure of the r.h.s. is far less
transparent in this case.

One also can approach to the Givental formula (\r{MF}) from another side. Namely, one can take
some known $\tau$-function, insert it into the l.h.s. and study the structure of the
differential operator at the r.h.s. The interesting example to deal with is provided by the
$\tau$-function for the Generalized Kontsevich Model
\cite{Kharchev:1992cp}.

\section{Acknowledgments}
The author is grateful to A. Mironov for careful reading the manuscript and useful discussions,
to A. Chervov, and V. Poberezhny  for
discussions and especially to A. Morozov for initiating this work and
careful support during the  work. I also acknowledge for the kind hospitality
Les Houches summer school and ICTP spring school on
superstrings and related matters. This work was partly supported by the $RFBR$ grant
00-02-16477, grant for support of young scientists 02-02-06504, 
Russian President's grant 00-15-99296 and grant $INTAS$-00-334.

\end{document}